\documentclass[prd,twocolumn,tightenlines,showpacs,nofootinbib,superscriptaddress,nobalancelastpage]{revtex4-1} 

\usepackage{graphics}     
\usepackage{graphicx}     
\usepackage{longtable}     
\usepackage{url}          
\usepackage{bm}           
\usepackage{natbib} 

\usepackage{textpos}
\usepackage{amsfonts,amsmath,amssymb,mathrsfs,bm,amsthm}
\usepackage{float}
\usepackage{booktabs}
\usepackage{array}
\usepackage{tabu}
\usepackage{dcolumn}
\usepackage{rotating}
\usepackage{ulem}
\usepackage{soul}

\newcommand{\RN}[1]{%
  \textup{\uppercase\expandafter{\romannumeral#1}}%
}

\usepackage{nicefrac}
\usepackage{amsthm}
\usepackage{color}
\usepackage{cancel}
\usepackage{appendix}
\usepackage{makecell}

\usepackage[colorlinks=true,linkcolor=black,citecolor=blue,urlcolor=blue,bookmarksopen]{hyperref}

\newcommand{\eref}[1]{(\ref{#1})}

\newcommand{\appropto}{\mathrel{\vcenter{
  \offinterlineskip\halign{\hfil$##$\cr
    \propto\cr\noalign{\kern2pt}\sim\cr\noalign{\kern-2pt}}}}}

\begin{document}
\title{Sensitivity of $^{229}$Th nuclear clock transition to variation of the fine-structure constant}

\date{\today}

\author{Pavel Fadeev}
\affiliation{Helmholtz Institute Mainz, Johannes Gutenberg University, 55099 Mainz, Germany}

\author{Julian C. Berengut}
\affiliation{School of Physics, University of New South Wales, Sydney, New South Wales 2052, Australia}

\author{Victor V.~Flambaum}
\affiliation{Helmholtz Institute Mainz, Johannes Gutenberg University, 55099 Mainz, Germany}
\affiliation{School of Physics, University of New South Wales, Sydney, New South Wales 2052, Australia}

\begin{abstract}
Peik and Tamm [Europhys. Lett. \textbf{61}, 181 (2003)] proposed a nuclear clock based on the isomeric transition between the ground state and the first excited state of thorium-229. This transition was recognized as a potentially sensitive probe of possible temporal variation of the fine-structure constant, $\alpha$. 
The sensitivity to such a variation can be determined from measurements of the  mean-square charge radius and quadrupole moment of the different isomers.
However, current measurements of the quadrupole moment are yet to achieve an accuracy high enough to resolve non-zero sensitivity.
Here we determine this sensitivity using existing measurements of the change in the mean-square charge radius, coupled with the ansatz of constant nuclear density. The enhancement factor for $\alpha$ variation is $K = -(0.82 \pm 0.25) \times 10^4$. For the current experimental limit,  $\delta\alpha/\alpha \lesssim 10^{-17}$ per year, 
the corresponding frequency shift is~ $\sim\!200$~Hz per year. This shift is six orders of magnitude larger than the projected accuracy of the nuclear clock, paving the way for increased accuracy of the determination of $\delta \alpha$ and interaction strength with low-mass scalar dark matter.
We verify that the constant-nuclear-density ansatz is supported by nuclear theory and propose how to verify it experimentally.
We also consider a possible effect of the octupole deformation on the sensitivity to $\alpha$ variation, and calculate the effects of $\alpha$ variation in a number of M\"ossbauer transitions.
\end{abstract}

\maketitle

\section{Introduction}
The first excited isomeric state of thorium-229, $^{\textrm{229m}}\textrm{Th}$, is a candidate for the first nuclear optical clock~\cite{PeikTamm2003}. This is due to the state's low excitation energy of several electron-volts \cite{Reich1990,Helmer1994,Beck2007,Beck2009} (the lowest of all known isomeric states) and long radiative lifetime of up to $10^4$ seconds \cite{Tkalya2015,Minkov2017}.
Several theoretical and experimental groups are making rapid progress toward using $^{\textrm{229m}}\textrm{Th}$ as a reference for a clock with unprecedented accuracy \cite{Porsev2010,Campbell2012,Wense2018,Thirolf2019,Tkalya2020,Nickerson2020,Porsev2010a,Porsev2010b,Dzyublik2020}. These papers proposed specific experimental schemes for the nuclear clocks and performed detailed studies of systematic effects such as the black body radiation shifts, effects of the ion  trapping fields in ion traps, and effects of the stray fields. The advantage of the nuclear clock in comparison with atomic clocks is that, due to the very small size of the nucleus and its shielding by the atomic electrons, it is insensitive to many systematic effects. For example, the nuclear polarizability and its contribution to the major systematic effect, the black body radiation shift, are 15 orders of magnitude smaller than in atomic transitions. Nuclear clocks may perform at a level of accuracy of $10^{-19}$~\cite{Campbell2012}, 1--2  orders of magnitude higher than the accuracy of the best existing atomic clocks. 

In a recent crucial step towards this goal, the transition was measured using spectroscopy of the internal conversion electrons emitted in flight during the decay of neutral $^{\textrm{229m}}\textrm{Th}$ atoms \cite{Seiferle2019}, yielding an excitation energy $E_{\text{is}} = 8.28\,(17)\,\mathrm{eV}$.  Another approach, using $\gamma$-ray spectroscopy at 29.2 keV, obtained $E_{\text{is}} =  8.30\,(92)\,\mathrm{eV}$~\cite{Masuda2019,Yamaguchi2019}. More recently, $E_\mathrm{is} =  8.10\,(17)\,\mathrm{eV}$ was reported~\cite{Sikorsky2020}.

The $^{\textrm{229m}}\textrm{Th}$ nuclear clock is expected to be a sensitive probe for time variation of the fundamental constants of nature \cite{Flambaum2006}. To avoid dependence on units we consider the effect of variation of the dimensionless fine-structure constant, $\alpha$,
related to the electromagnetic interaction~\cite{Flambaum2006,Auerbach,Litvinova09,Thirolf2019b,Julian2009,Julian2010}. 
Another dimensionless parameter, $m_q/\Lambda _{QCD}$, where $m_q$ is the quark mass and $\Lambda _{QCD}$ is the QCD scale,
is related to the strong interaction. The effect of $m_q/\Lambda _{QCD}$ variation on the $^{229}$Th transition has been estimated in Refs.~\cite{Flambaum2006,Xiao,Wiringa}. The high sensitivity to $\alpha$ comes about because the change in Coulomb energy between the isomers, which depends linearly on $\alpha$, is almost entirely cancelled by the nuclear force contribution which has only weak $\alpha$-dependence. This cancellation makes the energy of the transition $E_{\text{is}} =8$ eV  low compared to typical nuclear transitions, so any change in $\alpha$ and the Coulomb energy leads to a relative change several orders of magnitude larger in the energy of the transition $\Delta E_{\text{is}} /E_{\text{is}} $.

We also should note that the measurements of the variation of the fundamental constants does not require absolute frequency measurement.
All that is required is high stability of the ratio of two frequencies with a different dependence on the fundamental constants \cite{Dzuba1999PRL,Dzuba1999PRA}. For example, it may be the ratio of the 8-eV nuclear transition frequency to that of an atomic clock transition in the Th ion, as considered in Ref.~\cite{FlambaumPorsev2009}.

The change in the nuclear transition frequency, $f$, between the isomeric state and the ground state, $\delta f$, for a given change in the fine-structure constant, $\delta \alpha$, is \cite{Flambaum2006}
\begin{align}
   h \, \delta f=\Delta E_{\text{C}}\frac{\delta \alpha}{\alpha} \, ,
\end{align}
where $\Delta E_{\text{C}}$ is the
difference in Coulomb energy between the two isomers. The enhancement factor $K$ is defined by
\begin{align}
  \frac{\delta f}{f} =K \frac{\delta \alpha}{\alpha} \, ,
\end{align}
where $K = \Delta E_{\text{C}}/E_{\text{is}}$. Therefore, to find the sensitivity of $^{\textrm{229m}}\textrm{Th}$ transition to variation in $\alpha$, one needs to know $\Delta E_{\text{C}}$. 

The Coulomb energy $E_{\text{C}}$ depends on the shape of the nucleus. Unlike atomic systems, which are spherical due to the $1 / r$ potential from the pointlike nucleus ($r$ is the distance from the nucleus), nuclear systems can have deformed shapes as the potential originates from the nucleons themselves. 
Reference \cite{Julian2009} showed that, by modeling the nucleus as a prolate spheroid \cite{book88}, $\Delta E_{\text{C}}$ can be deduced from measurements of the change in nuclear charge radius and quadrupole moment between the isomeric and the ground states.
Using this model with measurements of nuclear parameters, the authors of \cite{Thielking2018} give a value of
\begin{align} \label{deltaEC} 
    \Delta E_{\text{C}}= -0.29\,(43) \,  \text{MeV} \, ,
\end{align}
where the dominant source of error is the uncertainty in measured quadrupole moments of the ground and the exited states. Such a  $\Delta E_{\text{C}}$ is consistent with a $K$ value anywhere between $0$ and $10^5$. This can be compared to a $K$ of about 
0.1--6 for current atomic clocks \cite{Dzuba1999PRL,Dzuba1999PRA,Huntemann2014,FlambaumDzuba2009,DFS2018,Safronova2019}.

In this paper we use the fact that the change in quadrupole moment is related to the change in charge radius to arrive at $\Delta E_{\text{C}}$ with errors consistent with a nonzero value, consequently giving a nonzero value for $K$. This relationship can be understood from the assumption of a constant charge density between isomers. We verify that this assumption gives a relation that is consistent with previous results from nuclear theory~\cite{Litvinova09}. 
We also test this assumption in several M\"ossbauer transitions, which we find have much lower sensitivities to $\alpha$ variation than the $^{229}\textrm{Th}$ transition.
Finally, following models that suggest the existence of an octupole deformation in $^{\textrm{229}}\textrm{Th}$, we use a more general treatment of a deformed nuclei. The results of the two models coincide within uncertainties.
\section{PROLATE SPHEROID MODEL}
We start by modeling the nucleus as a prolate spheroid with semiminor and semimajor axes $a$ and $c$. The volume $\left( 4 \pi /3 \right) R_0^3$ depends on $a$ and $c$ by
\begin{align} \label{constR0}
    a^2 c =R_0^3 \, .
\end{align}
The eccentricity $e$ is defined by 
\begin{align}
    e^2 = 1- \frac{a^2}{c^2}\,,
\end{align}
while the mean-square radius $\langle r^{2}\rangle$ and the quadrupole moment $Q_0$ are
\begin{align}  \label{r2}
    \langle r^{2}\rangle &= \frac{1}{5} \left( 2a^2 +c^2 \right) \, , \\
    Q_0 &= \frac{2}{5} \left( c^2 - a^2 \right) \, \nonumber.
\end{align}
The Coulomb energy can be written as a product of $ E_{\mathrm{C}}^{0}$, the Coulomb energy of an undeformed nucleus, and an anisotropy factor due to the deformation, $B_{\mathrm{C}}$ \cite{Carlson1961},
\begin{align} \label{Ec}
E_{\mathrm{C}} &= E_{\mathrm{C}}^{0} \, B_{\mathrm{C}} \, ,
\end{align}
where
\begin{align} \label{Eczero}
E_{\mathrm{C}}^{0} &=\frac{3}{5} \frac{q_e^{2} Z^{2}}{R_{0}} \, ,
\\  \label{Bcoul}
B_{\text{C}} &= \frac{(1-e^2)^{1/3}}{2e} \textrm{ln} \left( \frac{1+e}{1-e} \right)  \, .
\end{align}
Here $q_e$ is the electron charge and $Z$ is the number of protons.

In previous works~\cite{Julian2009}, $Q_0$ and $\langle r^{2}\rangle$ were treated as independent parameters. As such, calculation of $\Delta E_{\text{C}}$ involved derivatives of $E_{\text{C}}$ both by $Q_0$ and by $\langle r^{2}\rangle$:
\begin{align} \label{derivativeEc}
\Delta E_{\text{C}}=
\langle r^{2}\rangle \frac{\partial E_{\text{C}}}{\partial \langle r^{2}\rangle}
\frac{\Delta\langle r^{2}\rangle}{\langle r^{2}\rangle}+
Q_{0}\frac{\partial E_{\text{C}}}{\partial Q_{0}}
\frac{\Delta Q_0}{Q_{0}} \, .
\end{align}
With current experimental values $\langle r^2 \rangle = (5.76~\textrm{fm})^2$ and $Q_0 = 9.8(1)$~fm$^2$~\cite{note1}, Eqs. \eqref{Ec} and \eqref{derivativeEc} give
\begin{align} \label{delEXP}
    \Delta E_{\text{C}}=
-485\,\mathrm{MeV}
\frac{\Delta\langle r^{2}\rangle}{\langle r^{2}\rangle}
+11.6\,\mathrm{MeV}
\frac{\Delta Q_0}{Q_{0}} \, .
\end{align}
Substitution of measured changes in mean-square radius and quadrupole moment~\cite{Thielking2018}, $\Delta \langle r^2 \rangle = 0.012\,(2)$
~fm$^2$ and $\Delta Q_0/Q_0 = -0.01(4)$, gives the limit \eqref{deltaEC}.

\begin{figure} [tb]
\includegraphics[width=0.45\textwidth]
{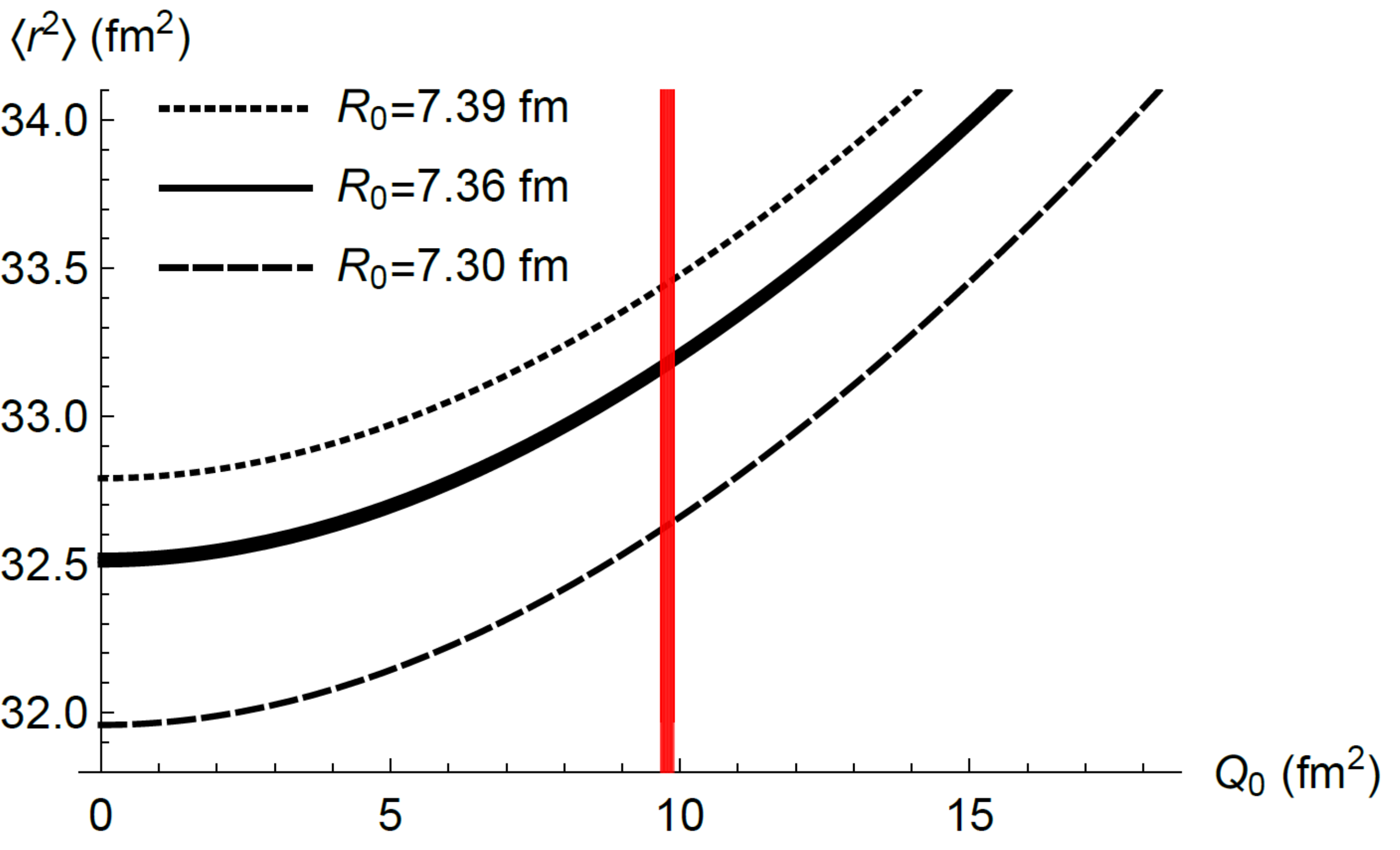}
\caption{Mean-square charge radius $\langle r^{2}\rangle$ as a function of intrinsic quadrupole moment $Q_0$ under the constant-volume ansatz for three volumes. The dashed lower curve corresponds to $R_0$ deduced from Hartree-Fock-Bogoliubov calculations using the SkM$^*$ functional, while the upper dotted curve is based on SIII functional (see Table~\ref{tab:nuccalc}). The middle curve, including errors, corresponds to $R_0 = 7.3615(16)$~fm deduced from the measurements by which  \eqref{deltaECnew} is obtained. The red line corresponds to the 1$\sigma$ experimental range of $Q_0$~\cite{note1}.}
\label{fig:r2vsQ0}
\end{figure}

Let us now consider the ansatz of constant charge density between isomers, equivalent to the ansatz of constant volume. That is, $R_0$ and hence $E_{\text{C}}^{0}$ are kept constant in the isomeric transition. Therefore, changes in $\langle r^2 \rangle$ and $Q_0$ are coupled by
\eqref{constR0} using \eqref{r2}. We show this dependence graphically in Fig.~\ref{fig:r2vsQ0}, and we can express it as
\begin{equation}
\label{eq:dQdr2}
\frac{dQ_0}{d\langle r^2 \rangle} = 1 + \frac{2\langle r^2 \rangle}{Q_0} = 7.8 \, ,
\end{equation}
where $7.8$ corresponds to the experimental values. Substitution of \eqref{eq:dQdr2} into \eqref{delEXP} gives us the following result:
\begin{equation}
\label{exprdE}
\Delta E_{\text{C}} = -180\,\mathrm{MeV}~\frac{\Delta \langle r^2 \rangle}{\langle r^2 \rangle} \, .
\end{equation}

The relation between changes in $\langle r^2 \rangle$ and $Q_0$ can also be obtained from nuclear calculations where the constant-density ansatz is not assumed. Results of the Hartree-Fock-Bogoliubov calculations of~\cite{Litvinova09} are summarized in Table~\ref{tab:nuccalc}. We extract $\Delta Q_0/\Delta\langle r^2 \rangle$ for two different energy functionals, SkM$^*$ and SIII, and for both protons and neutrons (for details see~\cite{Litvinova09}). In all cases the derivative is close to that predicted by the constant-density ansatz.

\begin{table}[tb]
\caption{Theoretical values of root-mean-square radius $r_{\text{rms}}$, $Q_0$, $\Delta Q_0$, and $\Delta r_{\text{rms}}$ calculated using Hartree-Fock-Bogoliubov approach with two energy functionals, SkM$^*$ and SIII.
In the fifth row we deduce the relationship between $\Delta Q_0$ and  $\Delta \langle r^2 \rangle$, which may be compared to the result of the constant density ansatz, ${dQ_0}/d{\langle r^2\rangle} = 7.8$.
In the last two rows we show the change in Coulomb energy from direct calculation and using \eqref{eq:dECdr2} with calculated values of charge radii.}
\centering
\begin{ruledtabular}
\begin{tabular}{lcccc}
 & \multicolumn{2}{c}{SkM$^*$} & \multicolumn{2}{c}{SIII} \\
 & $n$ & $p$ & $n$ & $p$ \\
\hline 
$r_{\text{rms}}$  (fm)$^\text{a}$
  & 5.8716 & 5.7078 & 5.8923 & 5.7769 \\
$Q_0$ (fm$^2$)$^\text{a}$
  & 9.2608 & 9.3717 & 9.0711 & 9.1643 \\
$\Delta Q_0$ (fm$^2$)$^\text{a}$
  & 0.2647 & 0.2756 & $-0.0516$ & $-0.0495$ \\
$\Delta r_{\text{rms}}$ (fm)$^\text{a}$
  & 0.0036 & 0.0039 & $-0.0005$ & $-0.0005$ \\
\hline
${\Delta Q_0}/\Delta{\langle r^2\rangle}$
  & 6.26 & 6.19 & 8.76 & 8.57 \\
\hline
$\Delta E_{\text{C}}$ (MeV)$^\text{b}$ & & $-0.307$ & & $0.001$ \\
$\Delta E_{\text{C}}$ (MeV)$^\text{a}$ & & $-0.287$ & & 0.036 \\
\end{tabular}
\end{ruledtabular}
\begin{flushleft}
$^\text{a}$ From Ref. \cite{Litvinova09}, Table II and Eq. \eqref{eq:dECdr2} for $\Delta E_{\text{C}}$. \\
$^\text{b}$ From Ref. \cite{Litvinova09}, Table I. \\
\end{flushleft}
\label{tab:nuccalc}
\end{table}

In addition to the results reproduced in Table~\ref{tab:nuccalc}, Ref.~\cite{Litvinova09} presents Hartree-Fock calculations (which do not include pairing) using the same functionals. For SkM$^*$, the Hartree-Fock calculations give the wrong sign for $\langle r^2 \rangle$, while for SIII the change between isomers is very small and susceptible to numerical noise. Nevertheless in both cases the Hartree-Fock calculations give reasonably close values for the derivative. 

For the Hartree-Fock-Bogoliubov calculations, the SkM$^*$  
better reproduces the measured energy interval and change in the nuclear radius between the isomers.
We take the average of the SkM$^*$ value $dQ_0/d\langle r^2 \rangle$ for protons and the experimental value from \eqref{eq:dQdr2} as our estimate of the derivative, and their difference as an estimate of the derivative's uncertainty, giving
${dQ_0}/d{\langle r^2\rangle} = 7.0\, (1.6)$. 
\footnote{Alternatively, we could use the average value between SkM* and SIII numbers for the derivatives $\Delta Q_0/\Delta r^2$ but the change in the result would be within the error bars.}
With this we write the change in Coulomb energy $\Delta E_{\text{C}}$ in terms of the change in mean-square radius at the physical point as
\begin{equation}
\label{eq:dECdr2}
\Delta E_{\text{C}} = -210\,(60)\,\mathrm{MeV}~\frac{\Delta \langle r^2 \rangle}{\langle r^2 \rangle} \, .
\end{equation}

The last row in Table~\ref{tab:nuccalc} lists the results of application of this formula to the nuclear calculations of $\Delta r_{\text{rms}}$ from~\cite{Litvinova09}.
Filling in the measured $\Delta \langle r^2 \rangle =~0.0105\,(13)~ \textrm{fm}^2$\,\cite{Safronova2018} and $\langle r^2 \rangle = (5.76~\textrm{fm})^2$ ~\cite{Angeli2013}, we obtain
\begin{align} \label{deltaECnew}
 \Delta E_{\text{C}} &= -0.067\,(19) \,\text{MeV} \, , \\
    K &= -0.82\,(25) \times 10^4 \, .
\end{align}
Since our model does not rely on the measured $\Delta Q_0$, which gives the biggest error in \eqref{deltaEC}, the result of  \eqref{deltaECnew} has smaller error than \eqref{deltaEC}.
We also predict $\Delta Q_0/Q_0 = 0.0075\,(20) \, \text{fm}^2$
which is within the experimental error of $\Delta Q_0/Q_0 = -0.01\,(4) \, \text{fm}^2$ presented in Ref. \cite{Thielking2018}.
\section{EFFECT OF OCTUPOLE DEFORMATION}
Nuclear calculations of  N. Minkov and A. P\'alffy suggest that the $^{\textrm{229}}\textrm{Th}$ nucleus has an octupole deformation \cite{Minkov2017,Minkov2019} (see also a recent experiment \cite{Chrishti}). They therefore describe the nucleus using a quadrupole-octupole model, obtaining a fair comparison to experimental results \cite{Minkov2017,Minkov2019}. This prompts us to include an octupole deformation in addition to the quadrupole deformation.

To facilitate this we describe the nucleus shape by its radius vector in axially symmetric spherical harmonics~\cite{Moller1995,Butler1996},
\begin{align}
\label{eq:radius_vector}
r(\theta) = R_{s}\left[ 1+ 
\sum_{n=1}^{N}\left(\beta_{n} Y_{n0}(\theta)\right)
\right] \, ,
\end{align}
where the coefficients $\beta_n$ are called deformation parameters and $N=3$ for the quadrupole-octupole model (pear shape). The length $R_s$ is defined by normalization of the volume to that of the undeformed nucleus,
\begin{align} \label{RsR0}
   \frac{2 \pi}{3} \int_0^{\pi}
     r^3(\theta) \sin\theta \,  d\theta \, 
    = \frac{4 \pi R_0^3}{3} \, .
\end{align}
The parameter $\beta_1$ is set such that the center of mass of the shape is at the origin of the coordinate system.

The mean-square radius and the intrinsic quadrupole moment of the nucleus are related to the deformation parameters $\beta_2$ and $\beta_3$ through $r(\theta)$ by
\begin{align} \label{rms}
\langle r^2 \rangle 
    &=  \int r^2(\theta) \rho(r) \, d^3r \, ,
\\ \label{quad}
Q_0 &= 2 \int r^2(\theta) P_2(\cos\theta) \rho(r) \, d^3r \, ,
\end{align}
where $\rho(r)$ is the charge density divided by the total charge. The factor 2 in \eqref{quad} is a matter of definition \cite{Jackson}, and fits with the special case of $Q_0$ in \eqref{r2}. 

\begin{figure} [tb]
\includegraphics[width=0.45\textwidth]
{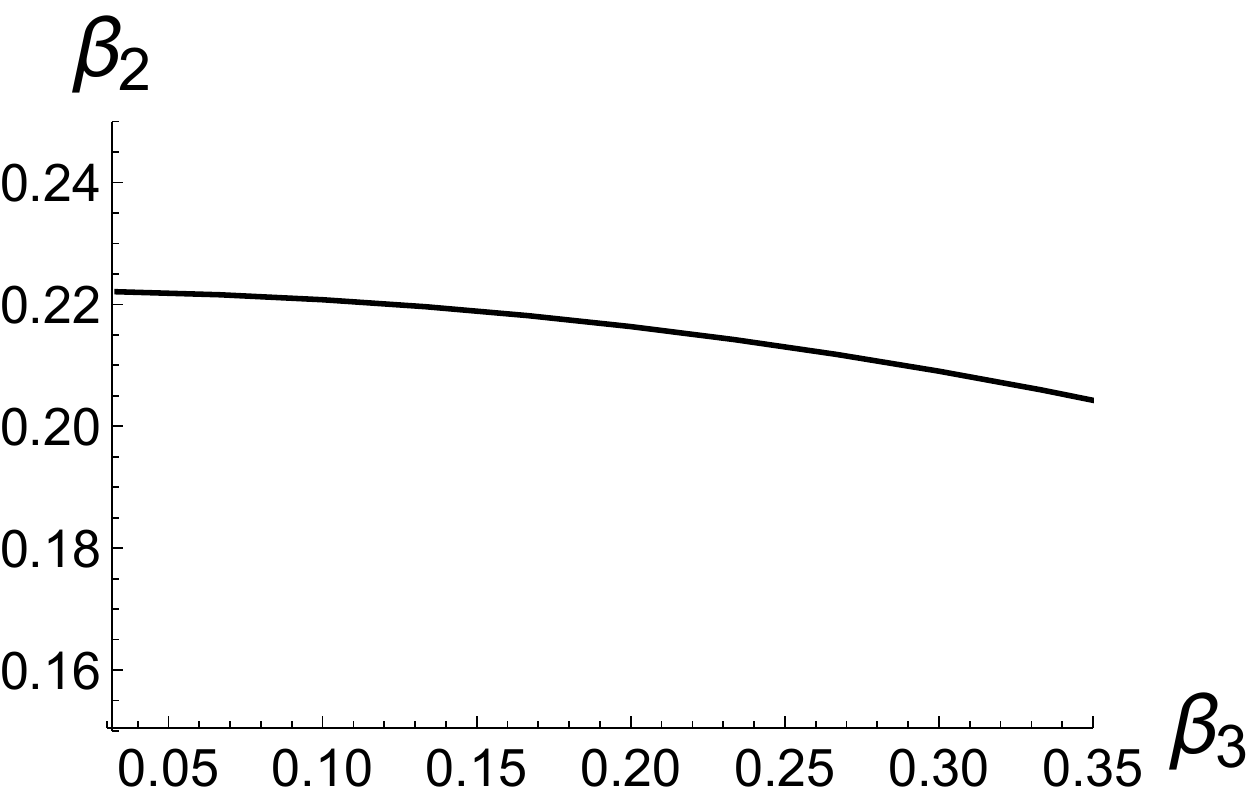}
\caption{Deformation parameter $\beta_2$ derived using \eqref{rms} and \eqref{quad} with experimental values of $Q_0 = 9.8\,\text{fm}^2$ and $\langle r^{2}\rangle=(5.76 \, \text{fm})^2$, as a function of $\beta_3$.}
\label{sensitivity}
\end{figure}

To determine $\beta_2$ for the pear shape, we solve \eqref{rms} and \eqref{quad} using the experimental values of $Q_0$ and $\langle r^{2} \rangle$. As the octupole moment of $^{\textrm{229}}\textrm{Th}$ has not yet been measured, we take $\beta_3=0.115$ from nuclear calculations~\cite{Minkov2017}. We arrive at $\beta_2=0.22$ and $R_s=7.3$~fm. This value of $\beta_2$ is fairly close to the theoretical prediction of \cite{Minkov2017}, $\beta_2=0.24$, and is not particularly sensitive to the chosen value of $\beta_3$ (see Fig.~\ref{sensitivity}).

In this model the anisotropy factor is \cite{book88}
\begin{equation}
\label{eq:BC_betas}
B_{\text {C}} = 1 - \frac{5}{4\pi} \sum_{n=2}^{\infty} \frac{n-1}{2 n+1} \beta_{n}^{2} 
+ \mathcal{O}(\beta_n^3) \ .
\end{equation}
Higher-order terms do not change our results within stated errors. With the aforementioned values for $\beta_2$ and $\beta_3$, we obtain for the constant-density ansatz (i.e., constant $E_C^0$),
\begin{align} \label{DEcbybeta}
\Delta E_{\text{C}} &= -76 \,\mathrm{MeV}\;\Delta\beta_2^2 -108 \,\mathrm{MeV}\;\Delta\beta_3^2 \\
\label{DEcbyr2andbeta}
&\approx -190 \,\mathrm{MeV}\;\frac{\Delta\langle r^2 \rangle}{\langle r^2 \rangle} -0.42 \,\mathrm{MeV}\; \frac{\Delta\beta_3^2}{\beta_3^2}\ .
\end{align}
Equation~\eref{DEcbyr2andbeta} is obtained by substituting~\eref{deltafuture}, and is in good agreement with \eref{exprdE}.
We see that the sensitivity of the nuclear clock to $\alpha$-variation does not depend strongly on the octupole moment.
\section{DISCUSSION}
The constant-volume ansatz used in the present work may be tested in experiments. This ansatz allows one to relate the change in nuclear quadrupole moment to the change in nuclear charge radius. Therefore, determination of $\Delta\langle r^2 \rangle$ by measuring the field isotope shift of atomic transitions, and extraction of $\Delta Q_0$ from the hyperfine structure or nuclear rotational bands, gives a measure of the change in the nuclear charge density.

A specific procedure can be encoded in the change in mean-square radius \cite{Ulm1986,Ahmad1988}
\begin{align}
\Delta \langle r^{2} \rangle
= \Delta \langle r^{2}\rangle_{\mathrm{sph}} +
  \Delta \langle r^{2}\rangle_{\mathrm{def}} \, .
\end{align}
Here the spherical part $\Delta \langle r^{2}\rangle _{\mathrm{sph}}$ describes the change in nuclear volume, i.e., volume contribution, and $\Delta \langle r^{2}\rangle _{\mathrm{def}}$ describes the deformation part assuming a constant volume, i.e., shape contribution. The latter can be expressed by deformation parameters \cite{Ulm1986,Ahmad1988,nucstr,Kalber1989}
\begin{align} \label{deltafuture}
\Delta \langle r^{2} \rangle = \Delta \langle r^{2} \rangle_{\mathrm{sph}} + \frac{5}{4 \pi} \langle r^{2} \rangle_\mathrm{{sph}} \left(\Delta \beta_{2}^2 + \Delta \beta_{3}^2+...\right),
  \end{align}
where $\langle r^{2}\rangle _{\mathrm{sph}}$ is the mean-square charge radius of the nucleus
assuming a spherical distribution. Equation ~\eqref{deltafuture} can be used in the future to test the volume-conservation hypothesis in isomers, once the $\Delta \beta$ is determined to higher accuracy. 

Using existing experimental data \cite{Thielking2018} we may conclude that the relative change in volume between $^{229}$Th isomers is less than a few parts per thousand, while the  calculations in \cite{Litvinova09} imply a fractional volume change of about $5 \times 10^{-4}$. This gives a quantitative evaluation of the constant-volume ansatz, which at times is used in the literature (see, e.g.,~\cite{Yordanov2016,Boucenna2003,Avgoulea2011}).

\begin{table*}[tb]
\caption{Sensitivity of M\"ossbauer transitions to variation of the fine structure constant. Coulomb energy shifts $\Delta E_{\text{C}}$ and enhancement factors $K$ are calculated using data from quadrupole moments~\cite{Stone2016} and isomeric shift measurements~\cite{Kalvius74}.
In columns 2 and 4 we use the constant density ansatz, where we have assumed a 25\% error from the ansatz, Eq.~\eqref{eq:dQdr2}, and 50\% error in the values of $\Delta \langle r^2 \rangle$ from  M\"ossbauer isomer shifts \cite{Kalvius74}.
The Eu isomers have more accurate values of $\Delta \langle r^2 \rangle$ taken from muonic x-ray and  M\"ossbauer data ~\cite{Tanaka1984Eu,Hess1969,Steichele1965}.
Columns 3 and 5 use experimental values of $\Delta Q_0$ from \cite{Stone2016} in the general formula~\eqref{derivativeEc}.
The ground-state $\langle r^2 \rangle$ values are taken from \cite{Angeli2013}.
$^{229}$Th results are shown for comparison (discussed in the main text).
}
\centering
\begin{ruledtabular}
\begin{tabular}{lcccc} 
 & \multicolumn{2}{c}{$\Delta E_{\text{C}}$ (MeV)} & \multicolumn{2}{c}{$|K|$} \\
 & constant density & general & constant density & general \\
\hline \\
$^{151}$Eu \, 22 keV
& $-0.099\,(51)$ & $-0.099\,(85)$ & $4.6\,(2.4)$ & $4.6\,(4.0)$ \\ \\
$^{153}$Eu 103 keV
& 0.32\,(18) & 0.02\,(15) & 3.1\,(1.8) & 0.2\,(1.5) \\ \\
$^{155}$Gd 105 keV
& 0.030\,(22) & 0.08\,(32) & 0.28\,(21) & 0.8\,(3.1) \\ \\
$^{157}$Gd
 \, 64 keV
& $-0.055\,(41)$ & $-0.06\,(21)$ & 0.86\,(63) & 0.9\,(3.3) \\ \\
$^{161}$Dy
 \, 75 keV 
 & $-0.031\,(23)$ & 0.29\,(55)  & 0.42\,(31) & 3.8\,(7.4) \\ \\
  $^{181}$Ta
\, \,  6 keV
 & 0.19\,(13) & 0.20\,(26) & 30\,(21) & 32\,(41) \\ \\
$^{243}$Am
\, 84 keV 
   & 0.23\,(17) & 0.45\,(75) & 2.8\,(2.0) & 5.4\,(9.0) \\ \\
$^{229}$Th
\, \, 8 eV
  & $-0.067\,(19)$ & $-0.26\,(39)$ &  $0.82\,(25) \, 10^4$ &  $3.1\,(4.8) \, 10^4$   \\ 
\end{tabular}
\end{ruledtabular}
\label{tab:moreelements}
\end{table*}

The sensitivity to potential variation of $\alpha$, i.e., the enhancement factor $K$, is three orders of magnitude larger than that of the most sensitive atomic clocks. 
For the present experimental bound  $\delta\alpha/\alpha \lesssim 10^{-17}$ per year, 
the frequency shift is up to $\sim\!200$~Hz per year. Since such a frequency shift is six orders of magnitude larger than the projected accuracy of the nuclear clock \cite{Campbell2012}, an unexplored range of $\delta \alpha$ may be tested.
As discussed in Refs. \cite{DM1,DM2,DM3}, the interaction between low-mass scalar dark matter and the electromagnetic field leads to oscillatory variation of $\alpha$. Therefore, the improvement in the sensitivity to $\alpha$ variation by six orders of magnitude afforded by such a clock should also lead to improved sensitivity in the search for low-mass scalar dark matter.

We should note a certain similarity between the research on $^{229}$Th isomeric transition and the very extensive experimental and theoretical studies  of isomeric (chemical) shifts in M\"ossbauer spectroscopy, which also involve effects of the change in the nuclear charge radius and electric quadrupole moment between the ground and excited nuclear states connected by a $\gamma$ transition. X-ray studies of muonic atoms are also able to deduce these nuclear properties.
Using the same technique as in $^{229}$Th we calculated the Coulomb energy difference $\Delta E_{\text{C}}$ and the relative sensitivity to $\alpha$ variation $K$ for nuclei where we have found sufficient experimental data. The results are presented in Table~\ref{tab:moreelements}. The enhancement factors $K$ for M\"ossbauer transitions ($K=\Delta E_C/E_{\text{is}} \sim $ 1 -- 30) are much smaller than $K$ for $^{229}$Th since the energy of M\"ossbauer transitions is much larger, $E \sim$ 5 -- 100 KeV. However, they are comparable or even bigger than $K\sim 0.1-6$ in atomic clocks. The energy resolution in M\"ossbauer $\gamma$ transitions may be as good as $10^{-18}$, see, e.g., the measurement of the gravitational redshift in Ref. \cite{GRS} where such a resolution was achieved after 5 days of measurements. This is even higher than that achieved recently in optical transitions, $10^{-17}$ -- $10^{-18}$. However, the authors of Ref. \cite{GRS} noted a problem with solid state effects which are difficult to control.

The results in Table~\ref{tab:moreelements} serve as a test of the constant density ansatz.
The predictions for $\Delta E_{\text{C}}$ using the constant-density model and using the more general formula, \eqref{derivativeEc}, with experimental data for both $\Delta \langle r^{2} \rangle$ and $\Delta Q_0$, agree within error bars. In these examples, using one of the values of $\Delta \langle r^2 \rangle$ or $\Delta Q_0$, the constant density ansatz reproduces the other value within error bars. This provides a check on the validity of the ansatz.

\bigskip
We thank Adriana P\'alffy, Anne Fabricant, Angela Moller, and Vadim Ksenofontov for their help.
The work was supported by the Australian Research Council Grants No. DP190100974 and DP20010015, the Gutenberg Fellowship, and the Alexander von Humboldt Foundation.

\bibliographystyle{prsty}

\end{document}